\documentclass[prl,twocolumn,floatix,amssymb,showpacs,amsmath,superscriptaddress]{revtex4-1}
\usepackage{graphicx}
\usepackage{epsfig}
\usepackage{amsmath,amssymb}
\begin{document}

\newcommand{\ket}[1]{\ensuremath{\left|{#1}\right\rangle}}
\newcommand{\bra}[1]{\ensuremath{\left\langle{#1}\right|}}
\newcommand{\quadr}[1]{\ensuremath{{\not}{#1}}}
\newcommand{\quadrd}[0]{\ensuremath{{\not}{\partial}}}
\newcommand{\slpar}{\partial\!\!\!/}
\newcommand{\gtrescero}{\gamma_{(3)}^0}
\newcommand{\gtresuno}{\gamma_{(3)}^1}
\newcommand{\gtresi}{\gamma_{(3)}^i}

\title{Quantum Simulation of Interacting Fermion Lattice Models in Trapped Ions}

\date{\today}

\author{J. Casanova}
\affiliation{Departamento de Qu\'{\i}mica F\'{\i}sica, Universidad del Pa\'{\i}s Vasco UPV/EHU, Apdo.\ 644, 48080 Bilbao, Spain}
\author{A. Mezzacapo}
\affiliation{Departamento de Qu\'{\i}mica F\'{\i}sica, Universidad del Pa\'{\i}s Vasco UPV/EHU, Apdo.\ 644, 48080 Bilbao, Spain}
\author{L. Lamata}
\affiliation{Departamento de Qu\'{\i}mica F\'{\i}sica, Universidad
del Pa\'{\i}s Vasco UPV/EHU, Apdo.\ 644,
48080 Bilbao, Spain}
\author{E. Solano}
\affiliation{Departamento de Qu\'{\i}mica F\'{\i}sica, Universidad del Pa\'{\i}s Vasco UPV/EHU, Apdo.\ 644, 48080 Bilbao, Spain}
\affiliation{IKERBASQUE, Basque Foundation for Science, Alameda Urquijo 36, 48011 Bilbao, Spain}

\begin{abstract}
We propose a method of simulating efficiently many-body interacting fermion lattice models in trapped ions, including highly nonlinear interactions in arbitrary spatial dimensions and for arbitrarily distant couplings. We map products of fermionic operators onto nonlocal spin operators and decompose the resulting dynamics in efficient steps with Trotter methods, yielding an overall protocol that employs only polynomial resources. The proposed scheme can be relevant in a variety of fields as condensed-matter or high-energy physics, where quantum simulations may solve problems intractable for classical computers.
\end{abstract}


\maketitle

Quantum simulations promise to revolutionize computing technologies and scientific research by allowing us to solve problems that are otherwise intractable with classical computers \cite{Feynman82,Lloyd96}. Several physical systems have been proposed in a quantum simulator: spin models \cite{Jane, Porras, Friedenauer08,Kim10}, quantum phase transitions \cite{Greiner02}, quantum chemistry \cite{Lanyon10}, anyonic statistics~\cite{Matthews}, many-body systems with Rydberg atoms~\cite{Weimar10}, and relativistic systems \cite{Lamata07, Gerritsma1,Casanova1,Gerritsma2,Casanova11,Lamata11,CasanovaQFT, Preskill2011}. Trapped ions offer one of the most promising platforms for quantum simulators~\cite{Schmidt-Kaler}, due to their high controllability, efficient initialization and measurement~\cite{WunderlichReview}. In the near future, experiments in quantum simulations should be able to solve problems intractable for classical computers, turning this technology into a remarkable tool for scientists and engineers. 

The numerical simulation of fermionic systems is, in general, a hard problem due to the huge increase of the Hilbert space dimension with the number of modes~\cite{Feynman82,Abrams97}. Using customized numerical methods as quantum Monte Carlo is not always possible due to the well-known sign problem~\cite{Troyer05, LohJR}.  Quantum simulations appear as a tool that will allow us to compute the time evolution of  free and interacting fermion lattice theories with minimal experimental resources.  This will be helpful in performing a wide range of condensed matter calculations, including those related with many-body interactions as the Kondo~\cite{Hewson97}, Fermi-Hubbard~\cite{Albers09}, or Fr\"ohlich~\cite{Mahan00} Hamiltonians.  Furthermore, quantum simulations will allow us to reproduce the complete dynamics of these systems, avoiding mean field approximations as Hartree-Fock to simplify nonlinear interactions~\cite{Schwabl05}.

In this Letter, we propose a method of realizing the quantum simulation of many-body fermionic lattice models for $N$ fermionic modes in trapped ions. Our method can be described in three steps. Firstly, we map the set of $N$ fermionic modes, via the Jordan-Wigner transformation~\cite{JordanWigner}, to a set of $N$ nonlocal spin operators. The second step consists in decomposing the total unitary evolution via Trotter expansion~\cite{Lloyd96, Nielsen00, Berry07} in terms of a product of exponentials associated to each nonlocal spin operator appearing in the Hamiltonian. Finally, we implement each of these exponentials in polynomial time on a set of $N$ two-level trapped ions with a reduced number of laser pulses~\cite{Mueller11,Barreiro11}. These three steps yield an efficient protocol that employs only polynomial resources. Our method can simulate highly nonlinear and long-range interactions in arbitrary spatial dimensions, applying the Jordan-Wigner transformation without the usual restriction of a reduced number of neighbors, and without the need of auxiliary virtual Majorana fermions~\cite{VerstraeteCirac05}. This is due to the fact that the dynamics associated with the nonlocal spin operators, containing a large number of Pauli matrices, can still be efficiently implemented. The proposed protocol opens the possibility to simulate a wide range of interesting condensed-matter and high-energy physics fermionic systems for a large number of particles. This includes the calculation of time evolutions and ground state computations, e.g. through adiabatic protocols~\cite{Farhi00}. For a number of particles above $\sim 30$, which is foreseeable in the near future, one could already simulate fermionic systems that are intractable for classical computers. 

We consider the quantum simulation of the dynamics associated with the general Hamiltonian 
\begin{equation}\label{themodel}
H = \sum_{n=2}^{\alpha}\bigg[\sum_{\ \ i_1...i_n =1}^{N} g_{i_1...i_n }  c_{i_1}  \cdots \ c_{i_n} +  {\rm H.c.}\bigg],
\end{equation}
where $c_{i_k}$ has to be chosen as one of the fermionic operators $b_{i_k}$, $b_{i_k}^{\dag}$,  that  obey the anticommutation rule $\{b_{i_k},b^\dag_{i_{k'}}\}=\delta_{i_k,i_{k'}}$, $N$ is the number of fermionic modes,  and $\alpha$ is the highest order of the many-body interaction. 

Our protocol consists of three steps, gathering techniques that have not been previously considered for quantum simulators of fermionic lattice models~\cite{SupplOnlMat}:

{\it i) Jordan-Wigner mapping.--}   This technique establishes a correspondence between a set of fermionic operators and a set of spin operators, transforming a local Hamiltonian of fermions onto a nonlocal Hamiltonian of spins. Only in one spatial dimension, and for couplings extended to a reduced number of neighbors, the correspondence is from a local model to a similar one~\cite{JordanWigner}. The operators $b_{i_k}, b^{\dag}_{i_k} $ can always be mapped to products of Pauli matrices, i.e., nonlocal spin operators, using the Jordan-Wigner transformation $b^{\dag}_k = I_N\otimes I_{N-1}\otimes...\otimes \sigma^{+}_{k}\otimes \sigma^{z}_{k-1}\otimes ...\otimes\sigma^{z}_1$, and  $b_{k}= (b^{\dag}_k)^{\dag}$.  

{\it ii) Trotter decomposition of the total Hamiltonian.--} Our second step consists in using standard Trotter techniques to decompose the total evolution operator in terms of a product of evolution operators associated to each nonlocal spin operator. We prove below that these evolution operators can be implemented efficiently.

{\it iii) Implementation of nonlocal spin operators in trapped ions.--} The exponentials of each nonlocal spin operator are efficiently implementable, given that their exponents consist of tensor products of $k$ Pauli matrices. Each of these exponentials can be implemented, for arbitrary $k$ and up to local rotations, with a M\o lmer-S\o rensen gate upon $k$ ions, one local gate upon one of the ions, and the inverse M\o lmer-S\o rensen gate~\cite{Mueller11}. This step can be summarized as
\begin{eqnarray}\label{ProductPauliMatrices}
{\cal U} &=& {\cal U}_{\rm MS}(-\pi/2, 0){\cal U}_{\sigma_z}(\phi){\cal U}_{\rm MS}(\pi/2,0)\nonumber\\
     &=& \exp \big[ i \phi \   \sigma^z_1 \otimes \sigma^x_2 \otimes \sigma^x_3 \otimes \cdot \cdot \cdot  \otimes \sigma^x_{k}\big],
\end{eqnarray}
where ${\cal U}_{\rm MS}(\theta,\phi)=\exp[-i\theta(\cos\phi S_x+\sin \phi S_y)^2/4]$, $S_{x,y}=\sum_{i=1}^k\sigma_i^{x,y}$ and ${\cal U}_{\sigma_z}(\phi)=\exp(i\phi'\sigma_1^z)$ for odd $k$, where $\phi'=\phi$ for $k=4n+1$, and $\phi'=-\phi$ for $k=4n-1$, with positive integer $n$. For even $k$, ${\cal U}_{\sigma_z}(\phi)$ would be substituted by ${\cal U}_{\sigma_y}(\phi)=\exp( i\phi'\sigma_1^y)$, where $\phi'=\phi$ for $k=4n$, and $\phi'=-\phi$ for $k=4n-2$, with positive integer $n$. 
In order to obtain directly a coupling composed of $\sigma^y$ matrices times a $\sigma^z$, one may apply a similar approach with different M\o lmer-S\o rensen gates according to
\begin{eqnarray}\label{ProductPauliMatricesBis}
{\cal U} &=& {\cal U}_{\rm MS}(-\pi/2, \pi/2){\cal U}_{\sigma}(\phi){\cal U}_{\rm MS}(\pi/2,\pi/2)\nonumber\\
     &=& \exp \big[ i \phi \   \sigma^z_1 \otimes \sigma^y_2 \otimes \sigma^y_3 \otimes \cdot \cdot \cdot \otimes  \sigma^y_{k}\big],
\end{eqnarray}
where the local ${\cal U}_{\sigma}(\phi)$ gate is $\exp(i\phi'\sigma^{z}_1)$ for odd $k$, where $\phi'=\phi$ for $k=4n+1$, and $\phi'=-\phi$ for $k=4n-1$, with positive integer $n$. For even $k$, the local gate is $\exp(i\phi'\sigma^{x}_1)$ where $\phi'=\phi$ for $k=4n-2$, and $\phi'=-\phi$ for $k=4n$, with positive integer $n$. 
Note that local rotations acting on each ion give rise to any tensor product of Pauli matrices $\sigma_{k}^{x, y, z}$.  The coupling constant in each nonlocal spin term of the simulated Hamiltonian~(\ref{themodel}) will be related to $\phi$ through $\phi=-gt$, where $g$ is a generic coupling strength and  $t$ is the corresponding gate time (for details, see~\cite{SupplOnlMat}).

The three steps of our protocol amounts to an efficient method for simulating fermionic models with long-range couplings in arbitrary dimensions with trapped ions.

Note that for bounded Hamiltonians, the Trotter expansion associated with the exponential of the polynomial sum of efficiently implementable nonlocal terms is also efficient~\cite{Lloyd96,Nielsen00,Berry07}, i.e., it only requires polynomial resources~\cite{SupplOnlMat}. This includes most fermionic models in condensed-matter and high-energy physics, some of which we consider below.

{\it Kondo model.--} The long debated Kondo Hamiltonian provides a variety of interesting features in different systems, as the minimum in the resistivity at low temperatures~\cite{Kondo1}. With the proposed method, we can simulate Kondo Hamiltonians~\cite{Hewson97}, modelling the interaction of a Fermi sea of electrons
with several magnetic impurities at positions $R_{j}$,
\begin{eqnarray}
H & = & \underset{p\sigma}{\sum}\epsilon_{p}b_{p\sigma}^{\dagger}b_{p\sigma}-J\underset{pp'j}{\sum}e{}^{iR_{j}\cdot(p-p')}[(b_{p\uparrow}^{\dagger}b_{p'\uparrow}-b_{p\downarrow}^{\dagger}b_{p'\downarrow})\sigma_{j}^{z}\nonumber\\&&+b_{p\uparrow}^{\dagger}b_{p'\downarrow}\sigma_{j}^{-}+b_{p\downarrow}^{\dagger}b_{p'\uparrow}\sigma_{j}^{+}].
\end{eqnarray}

 Here, $\sigma=\uparrow,\downarrow$ is the spin of the electron, $b_{p,p'\uparrow}(b^\dag_{p,p'\uparrow})$ is the annihilation(creation) operator for an electron with respective momentum $p$ or $p'$ and spin up, $\sigma_{j}^{+}(\sigma_{j}^{-})$ is the impurity raising(lowering) spin operator, $\epsilon_p$ is the energy of the kinetic electronic Hamiltonian and $J$ is the electron-impurity coupling. Notice that, e.g., the operators $b_{p\uparrow}^{\dagger}b_{p'\downarrow}\sigma_{j}^{-}$ can be now mapped to a sum of products of Pauli matrices, leading to an efficient implementation.

{\it Fermi-Hubbard model.--} The Fermi-Hubbard Hamiltonian~\cite{Albers09} takes into account a range of effects in condensed-matter physics, as the Mott transition, and is also believed to be relevant in high-$T_c$ superconductivity. It takes the form
\begin{equation}\label{exampleHubbardHamiltonian}
H=w\underset{\delta i\sigma}{\sum}b_{i\sigma}^{\text{\ensuremath{\dagger}}}b_{i+\delta\sigma}+U\underset{j}{\sum}b_{j\uparrow}^{\dagger}b_{j\uparrow}b_{j\downarrow}^{\dagger}b_{j\downarrow},
\end{equation}
where the first fermionic operator subindex refers to the lattice site and the second to the spin, $w$ is the hopping energy, $U$ is the onsite Coulomb repulsion and one usually makes the tight-binding approximation $\delta=\pm 1$. Notice that our method is general and extends the hopping terms to arbitrarily distant pairs of electrons. The last term contains the product of four fermionic operators, allowing for efficient implementation.

We could as well implement the coupling of arbitrary products of fermionic operators [similarly to Eq. (\ref{themodel})] to linear sum of bosonic operators,
\begin{eqnarray}\label{themodelWithBoson}
H   =  \sum_{n=2}^{\alpha}\bigg[\!\!\!\sum_{\ \ i_1... i_n =1}^{N} \!\!\!\!\!g_{i_1... i_n }  c_{i_1} \cdots \ c_{i_n}\sum_{j}g_j(a_j+a^\dag_j)+  {\rm H.c.}\bigg].\nonumber\\
\end{eqnarray}

The bosonic operators $a_j$ can be implemented with the vibrational modes of the ion chain. One would now consider the same gate sequence as in Eq. (\ref{ProductPauliMatrices}) but replacing ${\cal U}_{\sigma_z}(\phi)=\exp(i\phi' \sigma^z)$ with ${\cal U}_{\sigma_z,a}(\phi)=\exp[i\phi' \sigma^z\sum_jg_j(a_j+a^\dag_j)]$. The latter can be implemented by using a red and a blue sideband interactions for each of the $a_j$ modes in the context of standard trapped-ion technology~\cite{Leibfried03}.

{\it Fr\"ohlich model.--} The Fr\"ohlich Hamiltonian~\cite{Mahan00} models the interaction of electrons with phonons,

\begin{equation}
H=\underset{p}{\sum}\frac{p^{2}}{2m}b_{p}^{\dagger}b_{p}+\omega_{0}\underset{q}{\sum}a_{q}^{\dagger}a_{q}+\underset{qp}{\sum}M(q)b_{p+q}^{\dagger}b_{p}(a_{q}+a_{-q}^{\dagger}),
\end{equation}
with $M(q)$ being the electron-phonon coupling. Here, $b_{p}$ is the electron annihilation operator that destroys an electron with momentum $p$, $a_{q}$ is the phonon annihilation operator with momentum $q$, $\omega_{0}$ is the phonon frequency, and $m$ is the electron mass. The last term  contains the product $b_{p+q}^{\dagger}b_{p}(a_{q}+a_{-q}^{\dagger})$, whose dynamics can be implemented with our technique above according to Eq.~(\ref{themodelWithBoson}).  We can simulate this kind of Hamiltonians in order to recover the polaron physics, a critical open issue for the deep understanding of correlated electrons in solids~\cite{Frohlich}.

\begin{figure}[t] \centering
\includegraphics[width=1\linewidth]{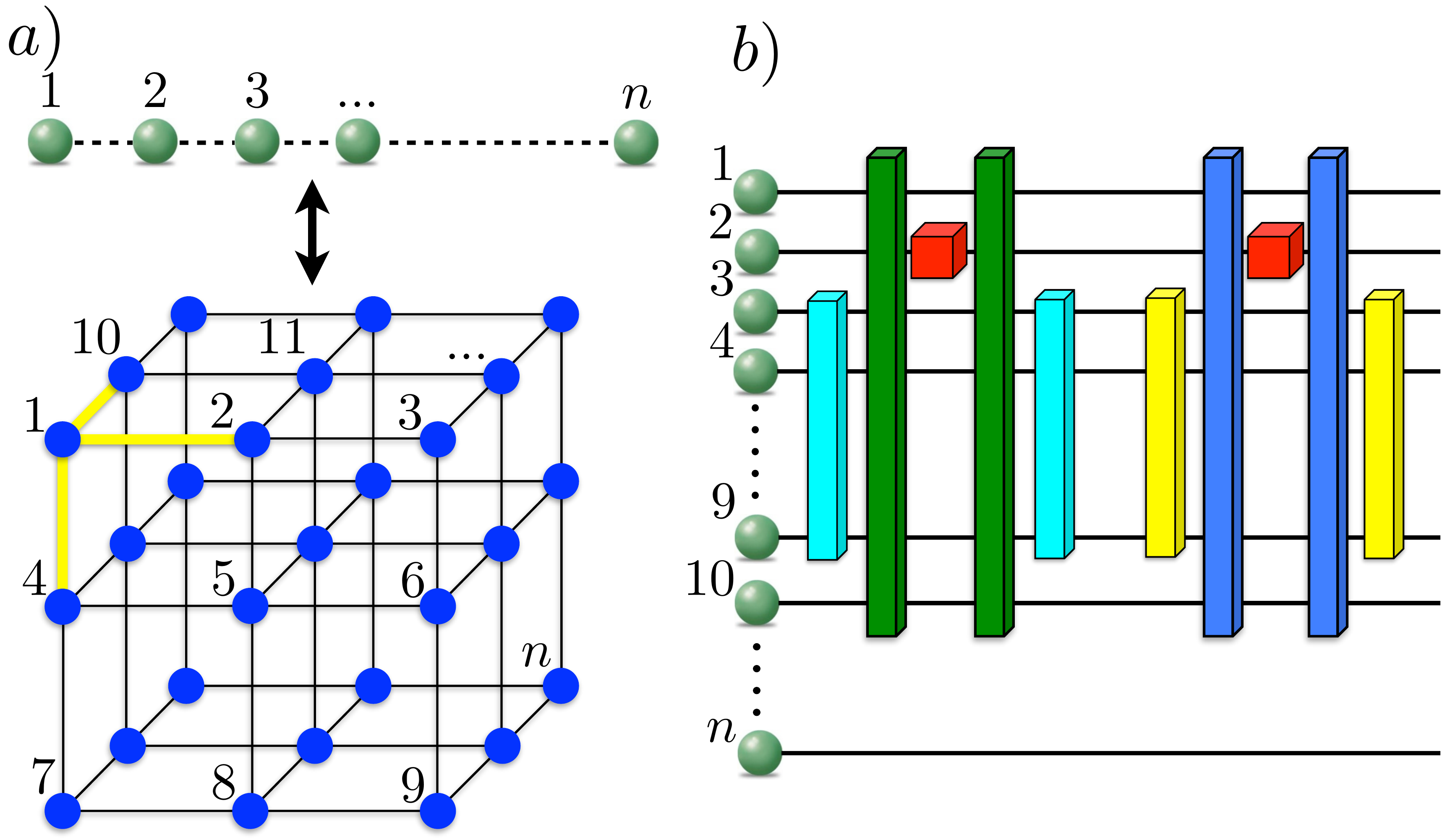}
\caption{(a) Mapping of a fermionic Hamiltonian onto an ion string. The couplings between fermions 1 and 4 (resp., 1 and 10) are nonlocal when applying the Jordan-Wigner transformation. (b) Efficient mapping of the tunneling coupling $b_1^\dag b_{10}+b_{10}^\dag b_1$ in trapped ions. This highly nonlocal coupling can be implemented with M\o lmer-S\o rensen gates (dark blue and green), local $\exp(i\phi'\sigma^y_2)$ gates (red), $\exp[\pm i(\pi/4)\sum_i\sigma_i^y]$ (yellow) and $\exp[\pm i(\pi/4)\sum_i\sigma_i^x]$ (cyan) gates.}
             \label{Fig1}
\end{figure}

One of the main appeals of our method is that the efficient encoding of fermionic models in a lattice with arbitrarily long-range couplings is feasible. This also means that we can apply the Jordan-Wigner transformation for two and three spatial dimensions, not just for one, without employing additional virtual Majorana fermions~\cite{VerstraeteCirac05}. All this is due to the fact that the fermionic operators are encoded in nonlocal spin operators whose dynamics are efficiently implementable. Thus, the mapped spin Hamiltonians are highly nonlocal but their evolution is efficiently realizable.
In order to show this, we plot in Fig.  \ref{Fig1} the mapping of a solid-state 3D fermionic system onto an ion string. As opposite to the nearest-neighbor tunneling coupling between fermions 1 and 2, which is local, the couplings between 1 and 4, and between 1 and 10, are nonlocal due to the Jordan-Wigner transformation (see yellow lines in Fig. \ref{Fig1}a). Nevertheless, we can implement them in an efficient way. In Fig.  \ref{Fig1}b we show the implementation of the tunneling coupling $b_1^\dag b_{10}+b_{10}^\dag b_1=\sigma^x_1\otimes\sigma^z_2\otimes\sigma^z_3\otimes...\otimes\sigma^z_9\otimes\sigma_{10}^x+\sigma^y_1\otimes\sigma^z_2\otimes\sigma^z_3\otimes...\otimes\sigma^z_9\otimes\sigma_{10}^y$ in trapped ions. This highly nonlocal coupling is a global unitary of $2^{10}\times 2^{10}$ dimensions. In the general case, it would require a number of elementary gates of $2^{20}\simeq 10^6$~\cite{Lloyd96}. With our method, the number of gates can be as small as 10 per Trotter step, consisting of M\o lmer-S\o rensen gates (dark blue and green), local $\exp(i\phi'\sigma^y_2)$ gates (red), $\exp[\pm i(\pi/4)\sum_i\sigma_i^y]$ (yellow) and $\exp[\pm i(\pi/4)\sum_i\sigma_i^x]$ (cyan) gates (see Fig. \ref{Fig1}b)~\cite{SupplOnlMat}. 

{\it Numerical simulations.--}
In order to compare the efficiency of the Trotter decomposition with the exact case, we have realized numerical simulations of the Fermi-Hubbard Hamiltonian, Eq. (\ref{exampleHubbardHamiltonian}), for different levels of Trotter expansion and for the exact diagonalization case. We have considered the case of three lattice sites, with six modes (two spins per site), to be simulated with six two-level trapped ions. The resulting Hamiltonian is
\begin{eqnarray}
&&H = w(b^\dag_{1\uparrow}b_{2\uparrow}+b^\dag_{1\downarrow}b_{2\downarrow}+b^\dag_{2\uparrow}b_{3\uparrow}+b^\dag_{2\downarrow}b_{3\downarrow}+{\rm H.c.})\nonumber\\
&&+U (b^\dag_{1\uparrow}b_{1\uparrow}b^\dag_{1\downarrow}b_{1\downarrow}
+b^\dag_{2\uparrow}b_{2\uparrow}b^\dag_{2\downarrow}b_{2\downarrow} 
+b^\dag_{3\uparrow}b_{3\uparrow}b^\dag_{3\downarrow}b_{3\downarrow}).\label{HubbardTrotter6ions}
\end{eqnarray}
Notice that the number of terms to be implemented scales linearly with the number of modes, $5N/2-4$ (11 in this case, for $N=6$). At the same time, the nonlocal gates upon several ions are efficiently implementable with few lasers, such that the number of gates in each term of the Hamiltonian is, in the worst case, linear in the number of modes, and in many cases just constant. In this particular example the total number of gates per Trotter step is of 33, i.e., on average 3 gates per Hamiltonian term, which is very efficient. This is due to the specific structure of this Hamiltonian, that avoids the need to apply additional local rotations.

\begin{figure}[t] \centering
\includegraphics[width=1\linewidth]{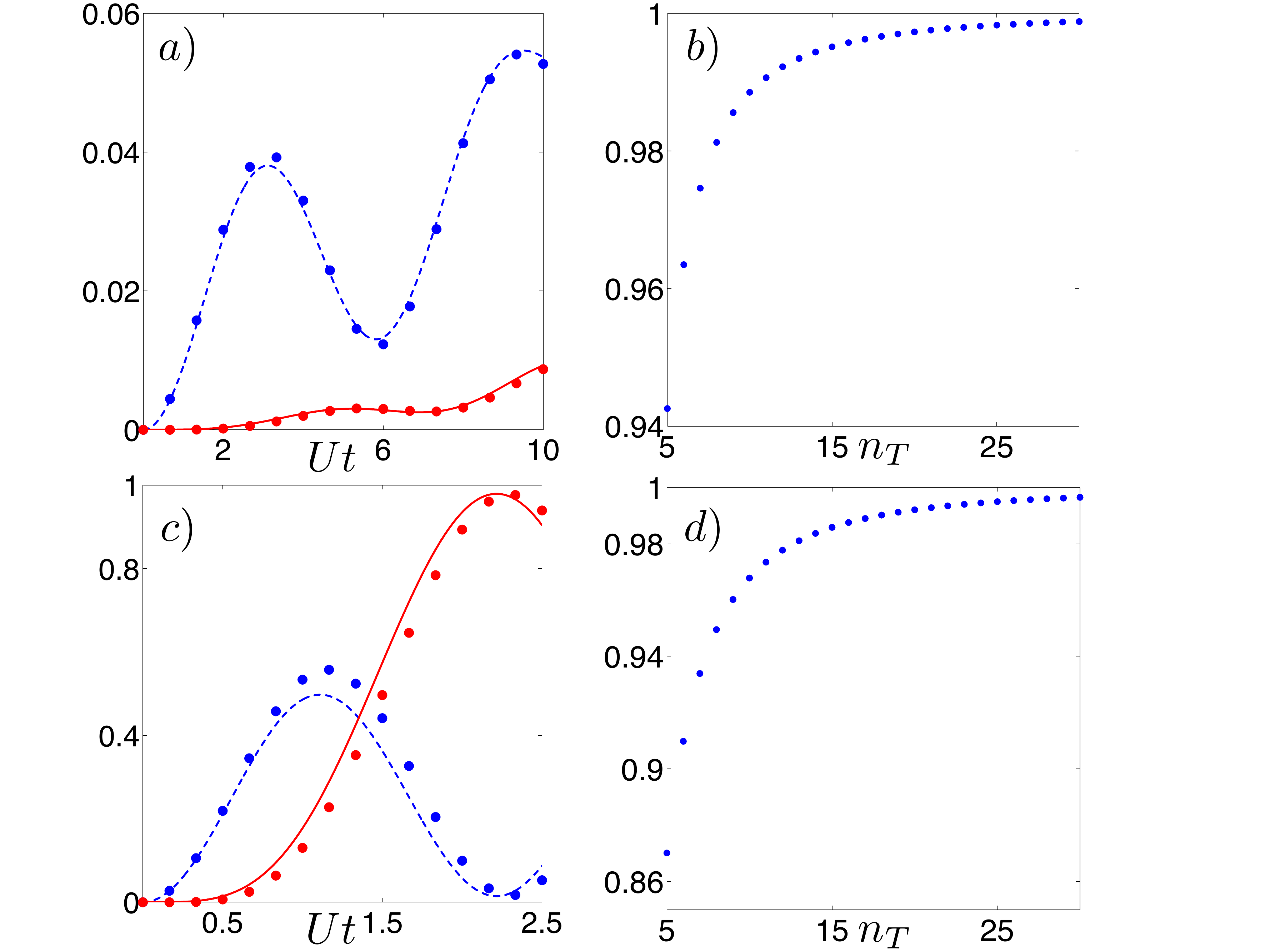}
\caption{(a) $\langle b^\dag_{2\downarrow}b_{2\downarrow}\rangle(t)$ (dashed, blue), and $\langle b^\dag_{3\uparrow}b_{3\uparrow}\rangle(t)$ (solid, red)  as a function of $Ut$, for a number of Trotter steps $n_T=15$, and (b) fidelity $|\langle \psi(t_F)|\psi(t_F)_T\rangle|^2$ as a function of $n_T$, for $U t_F=10$, where $|\psi(t)\rangle$ is the state evolved with exact diagonalization, and $|\psi(t)_T\rangle$ is the Trotter-evolved state, for $|\psi(0)\rangle=|\psi_T(0)\rangle= b^\dag_{1\uparrow}b^\dag_{1\downarrow}|0\rangle$, for $|w|/U=0.1$. (c) $\langle b^\dag_{2\downarrow}b_{2\downarrow}\rangle(t)$ (dashed, blue), and $\langle b^\dag_{3\uparrow}b_{3\uparrow}\rangle(t)$ (solid, red), for $n_T=15$, and (d) fidelity $|\langle \psi(t_F)|\psi(t_F)_T\rangle|^2$, for $U t_F=2.5$, where $|\psi(t)\rangle$ is the state evolved with exact diagonalization, and $|\psi(t)_T\rangle$ is the Trotter-evolved state, for $|\psi(0)\rangle=|\psi_T(0)\rangle= b^\dag_{1\uparrow}b^\dag_{1\downarrow}|0\rangle$, for $|w|/U=4$. In (a) and (c), the lines are obtained with exact diagonalization and the dots with Trotter expansion.}
             \label{Fig2}
\end{figure}

In Fig. \ref{Fig2}, we plot in (a) and (c) the average number of excitations for mode $b_{2\downarrow}$ (dashed, blue) and $b_{3\uparrow}$ (solid, red) for different parameters, showing the good convergence of the Trotter method. In (a) and (c), the lines are obtained with exact diagonalization and the dots with Trotter expansion. For additional figures with larger number of Trotter steps, see~\cite{SupplOnlMat}.  In (b) and (c) we include the fidelity  $|\langle \psi(t_F)|\psi(t_F)_T\rangle|^2$ as a function of  the number of Trotter steps $n_T$, where $|\psi(t)\rangle$ is the state evolved with exact diagonalization, and $|\psi(t)_T\rangle$ is the Trotter-evolved state, for $|\psi(0)\rangle=|\psi_T(0)\rangle= b^\dag_{1\uparrow}b^\dag_{1\downarrow}|0\rangle$, and for different parameters.
 Notice that the fidelity of the Trotter expansion goes to 1 with the number of Trotter steps $n_T$, and for $n_T=10$ it is 0.99 (b) or 0.97 (d).

With current technology, more than 100 gates have been realized in a single experiment~\cite{Lanyon11}. Indeed, without error correction, one would expect the realization of more than thousand gates in the near future~\cite{BlattPrivate}. This will allow, for $n_T=10$ Trotter steps, the implementation of hundreds of gates per step, giving us the possibility to simulate a wide variety of fermionic models.  In a possible experiment, one could consider, e.g., strings of Ca$^+$ ions controlled with lasers. The spin degrees of freedom can be encoded in long-lived electronic states of the ions~\cite{Gerritsma1,Gerritsma2,Barreiro11,Lanyon11,Friedenauer08,Kim10,Schmidt-Kaler}. Optimal state initialization via optical pumping and high-fidelity detection through resonance fluorescence can be easily performed. 

In conclusion, we have introduced a method for simulating efficiently, i.e. with polynomial resources, many-body fermionic lattice models in an ion string. These results may be relevant for quantum simulations of condensed-matter systems or high-energy physics in nonperturbative regimes.

We acknowledge discussions with R. Gerritsma and J. J. Garc\'{\i}a-Ripoll. The authors acknowledge funding from Basque Government BFI08.211 and IT472-10, EC Marie Curie IEF grant, Spanish MICINN FIS2009-12773-C02-01, UPV/EHU UFI 11/55, SOLID, CCQED, and PROMISCE European projects.

\section{Supplementary Material for ``Quantum Simulation of Interacting Fermion Lattice Models in Trapped Ions''}

\maketitle

\section{Implementation of fermion lattice models in trapped ions: Examples}

To illustrate our method, we propose several examples with increasing degree of complexity. In our first example, in order to implement the evolution associated with $H_1=g(b^\dag_{j+1}b_{j}+b^\dag_j b_{j+1})= g(I_N\otimes I_{N-1}\otimes...\otimes \sigma^{+}_{j+1}\otimes \sigma^{z}_{j}\sigma^{-}_{j}\otimes I_{j-1}\otimes...\otimes I_1+ I_N\otimes I_{N-1}\otimes...\otimes \sigma^{-}_{j+1}\otimes \sigma^{+}_j\sigma^{z}_{j}\otimes I_{j-1}\otimes...\otimes I_1)$ in terms of M\o lmer-S\o rensen gates, we first write $H_1$ as a sum of tensor products of Pauli matrices. This can always be done for Hermitian fermionic Hamiltonians: using $\sigma^\pm=(\sigma^x\pm i \sigma^y)/2$, we get $H_1=g(\sigma^x_{j+1}+i\sigma^y_{j+1})/2\otimes\sigma^z_j(\sigma^x_j-i\sigma^y_j)/2+(\sigma^x_{j+1}-i\sigma^y_{j+1})/2\otimes(\sigma^x_j+i\sigma^y_j)\sigma^z_j/2=-(g/2)(\sigma^x_{j+1}\otimes\sigma^x_j+\sigma^y_{j+1}\otimes\sigma^y_j)$. Once the fermionic Hamiltonian is in the form of sum of products of Pauli matrices, we use Trotter techniques to decompose the total evolution operator in product of exponentials associated with these products of Pauli matrices. Here, each of these exponentials is $\exp(i(g/2)t\sigma^x_{j+1}\otimes\sigma^x_j)$, and $\exp(i(g/2)t\sigma^y_{j+1}\otimes\sigma^y_j)$. The first one can be implemented using Eq. (2) in the article, with ${\cal U}_{\rm MS}(-\pi/2, 0)$, ${\cal U}_{\rm MS}(\pi/2,0)$ acting on ions $j$ and $j+1$, and ${\cal U}_{\sigma_y}(\phi'=-\phi=-gt/2)$ acting on ion $j+1$, plus a local rotation upon ion $j+1$ to change from the $z$ to the $x$ basis. Equivalently, for the second exponential, one can proceed similarly, substituting the M\o lmer-S\o rensen gates ${\cal U}_{\rm MS}(-\pi/2, 0)$, ${\cal U}_{\rm MS}(\pi/2,0)$ by ${\cal U}_{\rm MS}(-\pi/2, \pi/2)$, ${\cal U}_{\rm MS}(\pi/2,\pi/2)$, acting on ions $j$ and $j+1$, and ${\cal U}_{\sigma_x}(\phi'=\phi=gt/2)$ acting on ion $j+1$, plus a local rotation upon ion $j+1$ to change from the $z$ to the $y$ basis.

A second example describes the implementation of the nonlinear dynamics associated with $H_2=g(b^\dag_jb_jb^\dag_{j+1}b_{j+1})$, where $j$, $j+1$ may represent two different modes in the same site, or in different sites. The $j$ index labels the fermionic mode representing any degree of freedom (e.g., spatial, momentum, or spin). In this case, $H_2=g\sigma^+_j\sigma^-_j\otimes\sigma^+_{j+1}\sigma^-_{j+1}$, which can be written as a sum of products of Pauli matrices, obtaining $H_2=g(\sigma^z_j+I_j)\otimes(\sigma^z_{j+1}+I_{j+1})/4=g\sigma^z_j\otimes\sigma^z_{j+1}/4+g\sigma^z_j/4+g\sigma^z_{j+1}/4$. Finally, we apply Trotter methods to decompose ${\cal U}=\exp(-iH_2t)$ in three exponentials: $\exp(-igt\sigma^z_j\otimes\sigma^z_{j+1}/4)$ will be implemented, according to Eq. (2) in the article,  with ${\cal U}_{\rm MS}(-\pi/2, 0)$, ${\cal U}_{\rm MS}(\pi/2,0)$ acting on ions $j$ and $j+1$, and ${\cal U}_{\sigma_y}(\phi'=-\phi=gt/4)$ acting on ion $j$, plus a local rotation upon ion $j+1$ to change from the $x$ to the $z$ basis. The other two exponentials, $\exp(-igt\sigma^z_j/4)$ and $\exp(-igt\sigma^z_{j+1}/4)$ can be implemented easily by local gates.

As a third example, we consider the implementation of the evolution associated with a tunneling Hamiltonian,
\begin{eqnarray}
&&H_3=g(b_1^\dag b_{10}+b_{10}^\dag b_1)=g(\sigma^x_1\otimes\sigma^z_2\otimes\sigma^z_3\otimes...\otimes\sigma^z_9\otimes\sigma_{10}^x\nonumber\\&&+\sigma^y_1\otimes\sigma^z_2\otimes\sigma^z_3\otimes...\otimes\sigma^z_9\otimes\sigma_{10}^y) ,
\end{eqnarray}
 in trapped ions (see Fig. 1 in the article). Here, modes 1 and 10 represent nearest neighbours in a 3D lattice. Note that due to the standard mapping of Jordan-Wigner transformation in three dimensions, the effective spin Hamiltonian is highly nonlocal. We first decompose the evolution operator $\exp(-iH_3t)$ with a Trotter expansion in terms of the exponentials of each of the two terms in $H_3$, i.e., $\exp(-igt\sigma^x_1\otimes\sigma^z_2\otimes\sigma^z_3\otimes...\otimes\sigma^z_9\otimes\sigma_{10}^x)$ and $\exp(-igt\sigma^y_1\otimes\sigma^z_2\otimes\sigma^z_3\otimes...\otimes\sigma^z_9\otimes\sigma_{10}^y)$. In order to obtain the first of these operators, we apply the M\o lmer-S\o rensen gates ${\cal U}_{\rm MS}(-\pi/2, 0)$, ${\cal U}_{\rm MS}(\pi/2,0)$ to 
ions 1 to 10, applying in between of them the local ${\cal U}_{\sigma_y}(\phi'=-\phi=gt)$ gate upon the second ion. Notice that, despite the large number of ions, each of these gates requires just two lasers globally addressed upon all ions. We consider the second ion in this case because, according to Eq. (2) in the article, the resulting nonlocal spin operator  acts with a $\sigma^z$ operator upon the ion which was acted upon with ${\cal U}_{\sigma_y}(\phi'=-\phi=gt)$. In our case, we want it to be ion 2, given that in  $\exp(-igt\sigma^x_1\otimes\sigma^z_2\otimes\sigma^z_3\otimes...\otimes\sigma^z_9\otimes\sigma_{10}^x)$ ion 2, and not ion 1, is acted upon by $\sigma^z$ operator. Instead of ion 2, each of the ions 3-9 could also have been chosen for this purpose. In this way, we reduce the amount of local gates needed afterwards. Accordingly, we have so far that ${\cal U}_{\rm MS}(-\pi/2, 0){\cal U}_{\sigma_y}(\phi'=-\phi=gt){\cal U}_{\rm MS}(\pi/2,0)=\exp(-igt\sigma^x_1\otimes\sigma^z_2\otimes\sigma^x_3\otimes...\otimes\sigma^x_9\otimes\sigma_{10}^x)$. To obtain the desired exponential, we rotate qubits 3-9 with a global single qubit rotation to change the $x$ basis to the $z$ basis. In order to obtain the second exponential operator, a similar combination of gates should be employed, but in this case the specific M\o lmer-S\o rensen gates and single qubit gates would be different, as seen in Ref.~\cite{Mueller11b}.

\section{Analysis of the Trotter error}

In this Section we show that the resources needed in our protocol, including number of elementary gates and time scaling, are polynomial on the Trotter error, the total time simulated, and the total size of the system, in terms of the number of fermionic modes. 

The Trotter expansion is a useful tool to express the evolution operator of a Hamiltonian that can be written as a sum of efficiently-implementable Hamiltonians, in terms of a certain product of the operators associated to each of these individual Hamiltonians. More specifically, if the Hamiltonian H can be written as a sum of $m$ terms, where $m$ is polynomial in $N$, $H=\sum_{j=1}^m H_j$, then the standard Trotter expansion reads~\cite{Lloyd96b,Nielsen00b,Berry07b},
\begin{equation}
e^{-iHt}=(\prod_{j=1}^m e^{-iH_jt/n_T})^{n_T}+{\cal O}[1/n_T].\label{TrotterBasic}
\end{equation}
Accordingly, by making $n_T$ very large, the error can be made as small as possible.

There are more sophisticated, higher order expansions so called Lie-Trotter-Suzuki methods~\cite{Suzukiab,Suzukibb,Suzukicb},  that have a better scaling of the errors.  Here, we will focus on time-independent Hamiltonians whose evolution operator is expanded in terms of a $k$-th order  Lie-Trotter-Suzuki  integrator. We will follow the formalism and error analysis of Ref.~\cite{Berry07b}. In this reference it is shown that decompositions of ${\cal U}=\exp(-iHt)$, where $H=\sum_{j=1}^m H_j$, can be carried out in the general form
\begin{equation}
\tilde{\cal U}=\prod_{l=1}^{N_{\rm e}} e^{-iH_{j_l} t_l},
\end{equation}
with error $||{\cal U}-\tilde{\cal U}||<\epsilon$.
Here, the total number of gates needed, $N_{\rm e}$, scales in the error $\epsilon$, the total evolution time $t$, and the norm of the Hamiltonian $H$, $||H||$, where $||.||$ is the 2-norm, in the form~\cite{Berry07b}
\begin{equation}
N_{\rm e} \leq m5^{2k}(m||H||t)^{1+1/2k}/\epsilon^{1/2k},
\end{equation}
provided $\epsilon\leq 1\leq 2m5^{k-1}||H||t$.

\begin{figure}
  \centering
  \includegraphics[width=\linewidth]{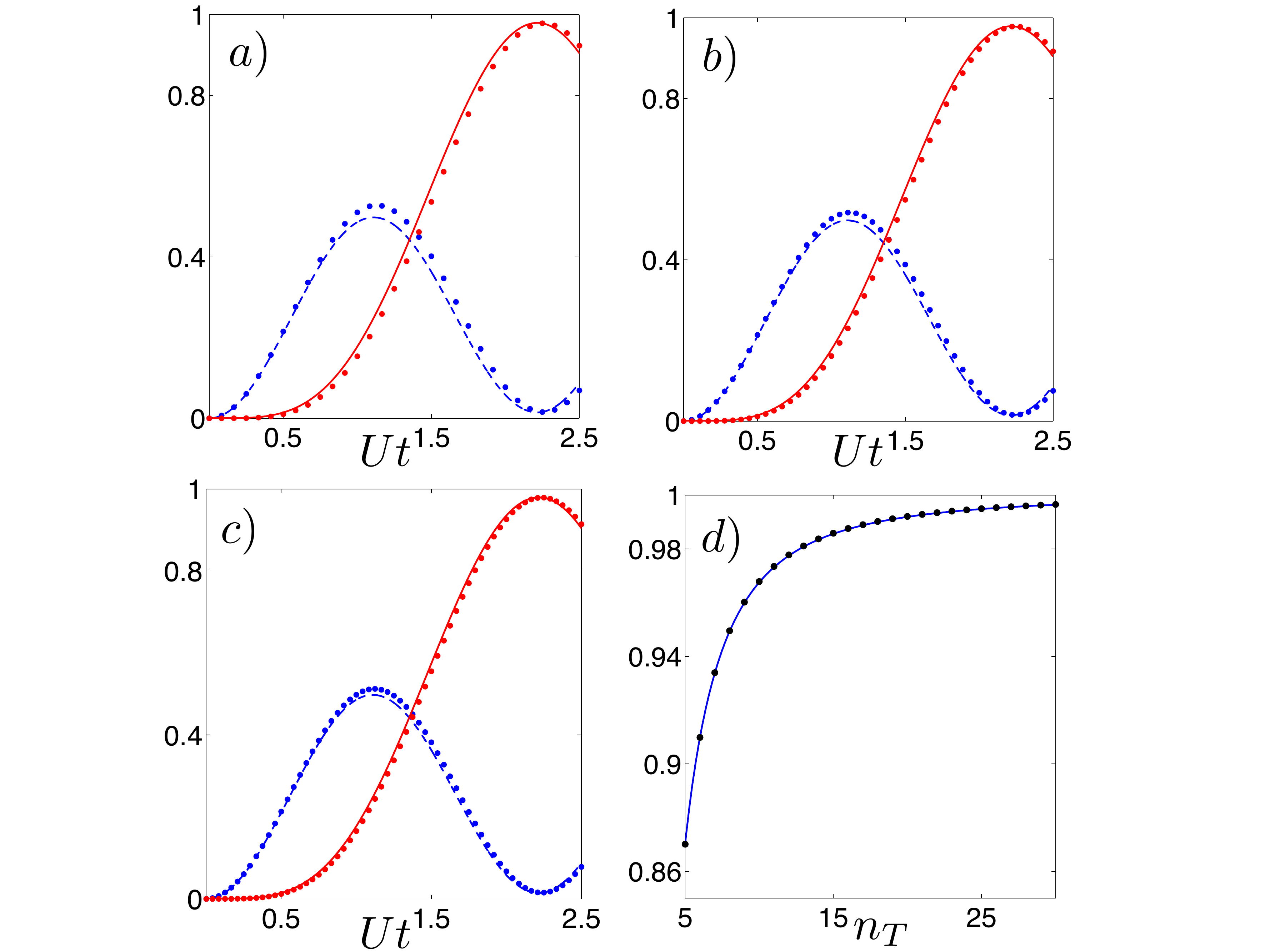}
  \caption{ We plot $\langle b^\dag_{2\downarrow}b_{2\downarrow}\rangle(t)$ (dashed, blue), and $\langle b^\dag_{3\uparrow}b_{3\uparrow}\rangle(t)$ (solid, red)  as a function of $Ut$, for $|\psi(0)\rangle=|\psi_T(0)\rangle= b^\dag_{1\uparrow}b^\dag_{1\downarrow}|0\rangle$,  $|w|/U=4$, and a number of Trotter steps $n_T=30$ (a), $45$ (b), and $60$ (c). In (a), (b), and (c) the lines are obtained with exact diagonalization and the dots are obtained with Trotter expansion. We also plot (d) the fidelity $|\langle \psi(t_F)|\psi(t_F)_T\rangle|^2$ as a function of $n_T$, for $Ut_F=2.5$, where $|\psi(t)\rangle$ is the state evolved with exact diagonalization, and $|\psi(t)_T\rangle$ is the Trotter-evolved state, for $|\psi(0)\rangle=|\psi_T(0)\rangle= b^\dag_{1\uparrow}b^\dag_{1\downarrow}|0\rangle$, and for $|w|/U=4$. We show the numerical results with Trotter (dots) and a fit to the function $1-C/n_T^2$ (line), where $C$ is a free parameter.}
  \label{fig:Supp}
\end{figure}

Notice that in all fermionic Hamiltonians we are considering, we have i) a polynomial number of nonlocal spin operators, i.e., $m$ is polynomial in $N$, the total number of fermionic modes. ii) each $H_j$ is always of the form of a product of arbitrary number of Pauli matrices times a coupling $h_j$, such that its norm $||H_j||=h_j$, given that the 2-norm of a product of arbitrary number of Pauli matrices is always 1. iii) the total norm of $H$ is bounded by $||H||\leq\sum_{j=1}^m||H_j||=\sum_{j=1}^m h_j\leq m h_j^{\rm max}$, where $h_j^{\rm max}$ is the maximum among all $h_j's$. Thus, $||H||$ is polynomial in $m$, and in consecuence, also in $N$.

Accordingly, we have shown that the scaling of the number of elementary gates needed in our expansion, is polynomial (more specifically, a power law) in $\epsilon$, $t$, and $N$, such that our method for implementing arbitrary fermionic Hamiltonians that occur in nature is efficient.  

\section{Final Remarks}

We plot Fig.  \ref{fig:Supp}a,b,c in order to analyze the convergence of Trotter methods to the exact diagonalization case when increasing the number of Trotter steps $n_T$, and comparing with Fig. 2 in the article (for which $n_T=15$). These three figures clearly show the fast convergence for a linear increase in $n_T$. 

In Fig. \ref{fig:Supp}d we plot the fidelity $|\langle \psi(t_F)|\psi(t_F)_T\rangle|^2$ as a function of $n_T$, for $Ut_F=2.5$, where $|\psi(t)\rangle$ is the state evolved with exact diagonalization, and $|\psi(t)_T\rangle$ is the Trotter-evolved state, for $|\psi(0)\rangle=|\psi_T(0)\rangle= b^\dag_{1\uparrow}b^\dag_{1\downarrow}|0\rangle$, and for $|w|/U=4$. We show the numerical results with Trotter (dots) and a fit to the function $1-C/n_T^2$ (line), where $C$ is a free parameter. This curve has a perfect agreement with the Trotter numerics. Thus, the error goes to zero polynomially in $n_T$, as expected.

\end{document}